\documentclass[11pt,prd,superscriptaddress,preprintnumbers,amsmath,amssymb,nofootinbib]{revtex4}
\usepackage{epsfig}  
\usepackage{graphicx}
\usepackage{hyperref}
\usepackage{color}
\usepackage{float}
\usepackage{amsfonts}
\usepackage{amsmath}
\usepackage{slashed}

\large

\begin{document} 
\title{Probing new physics through $B^*_s \rightarrow \mu^+ \mu^-$ decay}

\author{Dinesh Kumar}
\email{dinesh@uniraj.ac.in}
\affiliation{Department of Physics, University of Rajasthan, Jaipur 302004, India}

\author{Jyoti Saini}
\email{saini.1@iitj.ac.in }
\affiliation{Indian Institute of Technology Jodhpur, Jodhpur 342011, India}

\author{Shireen Gangal}
\email{shireen.gangal@gmail.com}
\affiliation{Center for Neutrino Physics, Department of Physics, Virginia Tech, Blacksburg, VA 24061, USA}

\author{Sanjeeda Bharati Das }
\email{sanjeedabd0194@gmail.com}
\affiliation{Indian Institute of Technology Jodhpur, Jodhpur 342011, India}
\affiliation{Department of Physics, Ramanujan Junior College, Nagaon 782001, India}

\date{\today} 

\preprint{}

\begin{abstract}
We perform a model independent analysis of new physics in $B^*_s \rightarrow \mu^+ \mu^-$ decay. We intend to identify new physics operator(s) which can provide large enhancement in the branching ratio of $B^*_s \rightarrow \mu^+ \mu^-$ above its standard model prediction. For this, we consider new physics in the form of vector, axial-vector, scalar and pseudoscalar operators. We find that scalar and pseudoscalar operators do not contribute to the branching ratio of $B^*_s \rightarrow \mu^+ \mu^-$. We perform a global fit to all relevant $b \to s \mu^+ \mu^-$ data for different new physics scenarios. For each of these scenarios, we predict $Br(B^*_s \rightarrow \mu^+ \mu^-)$. We find that a significant 
 enhancement in  $Br(B^*_s \rightarrow \mu^+ \mu^-)$ is not allowed by any of these new physics operators. In fact, for all new physics scenarios providing a good fit to the data, the branching ratio of $B^*_s \rightarrow \mu^+ \mu^-$ is suppressed as compared to the SM value. Hence  the present $b \to s \mu^+ \mu^-$ data indicates that the future measurements of $Br(B^*_s \rightarrow \mu^+ \mu^-)$ is expected to be suppressed in comparison to the standard model prediction.
\end{abstract}

\maketitle 

\section{Introduction} 

The Standard Model (SM) of particle physics is in good agreement with almost all observed experimental data till date but still it has various limitations such as it cannot account for the observed matter  dominance over anti-matter in our universe, existence of dark matter and dark energy cannot be explained within the SM.  Therefore, it is natural to look for physics beyond the SM. Recently, there have been several measurements in the B meson sector which do not agree with the predictions of SM. Although these measurement are not statistically significant, they can still provide signatures of physics beyond the SM. 
Many such measurements are in decays induced by the flavor changing neutral current quark level transition $ b \rightarrow s\, l^+ \, l^-$.

The measurement of $R_K \equiv  \Gamma(B^+ \to K^+ \,\mu^+\,\mu^-)/\Gamma(B^+ \to K^+\,e^+\,e^-)$ \cite{rk} by the LHCb collaboration, performed in the low dilepton invariant mass-squared $q^2$ range ($1.0 \le q^2 \le 6.0 \, {\rm GeV}^2$), deviates from the SM prediction of $\simeq 1$ \cite{Hiller:2003js, Bordone:2016gaq}, by 2.6 $\sigma$. Recently, the measurement of $R_{K^*} \equiv \Gamma (B^0 \to K^{*0} \mu^+\mu^-)/\Gamma(B^0 \to K^{*0} e^+ e^-)$ in the low ($0.45 \le q^2 \le 1.1 \, {\rm GeV}^2$)  and central ($1.1 \le q^2 \le 6.0 \, {\rm GeV}^2$) $q^2$ bins \cite{rkstar} enforces the lepton flavor universality violation in the $b \to s l^+ l^-$ sector. 
$R_{K^*}$ measurements differ from the SM prediction of  $\simeq 1$ \cite{Hiller:2003js, Bordone:2016gaq}
 by 2.2-2.4$\sigma$ and 2.4-2.5$\sigma$, in the low and central $q^2$ regions, respectively. This can be accounted by the presence of new physics in $b \to s \mu^+ \mu^-$ or $b \to s e^+ e^-$ sector. 
However, apart from the measurements of $R_K$ and $R_{K^*}$, there are other measurements which show disagreement with the SM predictions. Measurement of some of the angular observables in $B \to K \mu^+ \mu^-$ \cite{Kstarlhcb1,Kstarlhcb2,KstarBelle} disagrees with the SM predictions \cite{sm-angular}. In particular, $P'_5$ in the 4.3-8.68 $q^2$-bin disagress with the SM at the level of 4$\sigma$. This disagreement is further supported by the recent measurements by ATLAS \cite{kstaratlas} and CMS \cite{kstarcms}. Apart from the angular observables, there is tension in the branching ratio of $B_s \to \phi \mu^+ \mu^-$ as well \cite{bsphilhc1,bsphilhc2}. All of these discrepancies are related to the  $b \to s \, \mu^+ \, \mu^-$ sector. Hence it is quite natural to account for all of these anomalies by assuming new physics in $b \to s \mu^+ \mu^-$ transition only. This is further supported by the recent global fits \cite{Capdevila:2017bsm}. 

These anomalies are indications of new physics which one needs to explore. 
In \cite{Alok:2010zd,Alok:2011gv}, new physics in $b \to s \, \mu^+ \, \mu^-$  decays were analysed 
in a model independent way by making use of an 
effective Hamiltonian with all possible Lorentz structures. It
was found that any large effects in  $b \to s \, \mu^+ \, \mu^-$ sector, in particular decays like $B \to K^* \mu^+ \mu^-$ and  $B_s \to \phi \mu^+ \mu^-$, can only be due to new physics vector ($V$) and axial-vector operators ($A$). Scalar  ($S$) and pseudoscalar ($P$) operators are insignificant for these decays.  
This fact is corraborated by several global fits using model independent analysis. These fits suggest various new physics solutions to explain anomalies in the $b \to s \mu^+ \mu^-$ decay and they are mainly in the form of $V$ and $A$ operators \cite{Descotes-Genon:2013wba,Altmannshofer:2014rta,Altmannshofer:2015sma,Hurth:2016fbr,Capdevila:2016ivx,Altmannshofer:2017fio,Alok:2017jgr,Capdevila:2017bsm,Alok:2017sui}. These new physics operators could affect other observables related to $b \rightarrow s \mu^+ \mu^-$ transitions as well. In order to discriminate between various solutions and pin down the type of new physics responsible for all anomalies in the decays included by $b \to s \mu^+ \mu^-$ transition, one should look for alternative observables. Also,  it would be desirable to have an access to observables which are theoretically clean.

The purely leptonic decay of $B_s^*$ meson is such a decay channel \cite{Grinstein:2015aua}. Its sensitivity to new physics is quite complementary to that of $B_s \to \mu^- \mu^-$ as  $Br(B_s^* \to \mu^- \mu^-)$  is sensitive to  different combinations of Wilson coefficients. Also, $Br(B_s^* \to \mu^- \mu^-)$ is  not chirally suppressed. Further, this decay is theoretically
very clean as the amplitude depends only upon
decay constants which are  accurately determined in the
lattice QCD and the invariant mass of the process, $q^2=m_{B^*_s}^2=28$ $\rm GeV^2$, is well
above the charmonium states. This enables the application of an
operator-product expansion for the nonlocal contributions
through the quark-hadron duality. Therefore
the $B_s^* \to \mu^- \mu^-$  decay rate can be accurately predicted in
the standard model provided the $B^*_s$ decay width is well known. However this width is neither measured experimentally not accurately determined theoretically. The  $B^*_s$ decay width is the only hindrance  in the clean determination of the branching ratio. In future, this situation can improve owing to lattice QCD calculations. Using $\Gamma \sim 0.1$ KeV,  the branching
fraction for this process is predicted to be $\sim 10^{-11}$ \cite{Grinstein:2015aua}. This can be  within
reach of next run of LHC.  

The impact of $B^*_s \to \mu ^+ \mu ^-$  on $B_s \to \mu ^+ \mu ^-$ was studied in \cite{Xu:2015eev}. In \cite{Sahoo:2016edx} this decay is investigated in scalar leptoquark and family non-universal $Z^{'}$ models. It was shown that the scalar leptoquark model can provide significant enhancement in the branching ratio of $B_s^* \rightarrow \mu^+ \mu^-$ whereas in  $Z^{'}$ model, large enhancement is not possible.

In this work we perform a model independent analysis of $B_s^* \rightarrow \mu^+ \mu^-$ decay by considering new physics in the form of $V$, $A$, $S$ and $P$ operators. We do not consider new physics tensor operators as it is very difficult to construct a new physics model leading to tensor operators. We find that $S$ and $P$ operators do not contribute to the branching ratio of $B_s^* \rightarrow \mu^+ \mu^-$. We intend to identify the new physics interactions which can provide large enhancement in the branching ratio of $B_s^* \rightarrow \mu^+ \mu^-$. Also, it would be interesting to see whether $Br(B_s^* \rightarrow \mu^+ \mu^-)$  can discriminate between various new physics solutions which provide a good fit to the $b \to s \mu^+ \mu^-$  data.  We first perform a global fit to all relevant $b \to s \mu^+ \mu^-$  data and identify various new physics solutions. For each of these solutions, we obtain predictions for branching ratio of $B_s^* \rightarrow \mu^+ \mu^-$. We find that a large enhancement in  $Br(B^*_s \rightarrow \mu^+ \mu^-)$ is not possible due to any of these new physics scenarios. 

The paper is organized as follows. In Section II, we discuss $B^{*}_{s} \to \mu^+\mu^-$ decay within SM and in the presence of new physics operators. In Section III, we discuss the methodology for the $\chi^2$ fit. The results are presented in Section IV. Finally we provide conclusions in Section V.

\section{$B^{*}_{s} \to \mu^+\mu^-$ decay }
The $B_s^*$,  is a vector meson, with the same quark content as the $B_s$ meson and can be used as a complementary probe to study semi-leptonic B decays. The branching fraction of  $B_s^* \to \mu^+\mu^-$ can  be precisely measured by the end of Run III of the LHC. A detailed calculation of the SM decay rate for the $B_s^* \rightarrow \mu^+ \mu^-$ process can be found in Ref. \cite{Grinstein:2015aua}. In this section we sketch the calculation, in brief, by using the effective Hamiltonian for the process $B^{*}_{s} \to \mu^+\mu^-$ in the SM and obtain the decay rate and branching ratio. We then explore new physics contributions to this process in a model-independent way by adding $V$, $A$, $S$ and $P$ operators to the SM effective Hamiltonian and calculate the decay rate. 

\subsection{Branching ratio of $B^{*}_{s} \to \mu^+\mu^-$ in the SM}
\label{subsecA}

The effective Hamiltonian for the quark level transition $ b\to s \mu^+ \mu^-$ within the SM is given by
\begin{align} \nonumber
\mathcal{H}^{SM} &= -\frac{ 4 G_F}{\sqrt{2} \pi} V_{ts}^* V_{tb} \bigg[ \sum_{i=1}^{6}C_i(\mu) \mathcal{O}_i(\mu) + C_7\frac{e}{16 \pi^2}[\overline{s} \sigma_{\mu \nu}(m_s P_L + m_b P_R)b] F^{\mu \nu} \\& +C_9 \frac{\alpha_{em}}{4 \pi} (\overline{s} \gamma^{\mu} P_L b)(\overline{\mu} \gamma_{\mu} \mu) + C_{10 } \frac{\alpha_{em}}{4 \pi} (\overline{s} \gamma^{\mu} P_L b)(\overline{\mu} \gamma_{\mu} \gamma_{5} \mu) \bigg],
\end{align}
where $G_F$ is the Fermi constant, $V_{ij}$ are elements of the Cabibbo-Kobayashi-Maskawa (CKM) matrix and $P_{L,R} = (1 \mp \gamma_{5})/2$. The short-distance structure of the $b \rightarrow s$ transition is contained in the SM Wilson Coefficients $C_i$'s of the respective operators $ \mathcal{O}_i$'s where $O_{9,10}$ are the semi-leptonic operators and $O_{7}$ is the electric dipole operator.  The effect of the operators $\mathcal{O}_i,~ i=1-6,8$ can be included in the effective Wilson Coefficients by redefining $C_7(\mu) \rightarrow C_7^{\mathrm{eff}}(\mu,q^2)$ and $C_9(\mu) \rightarrow C_9^{\mathrm{eff}}(\mu,q^2)$.

The matrix elements of the operators $O_{7,9,10}$ can be related to the decay constant, $f_{B_s^*}$ of $B_s^*$ meson as follows  \cite{Grinstein:2015aua}
\begin{align} \nonumber
\langle 0 | \overline{s}\gamma^{\mu}b B^{*}_{s}(p_{B^{*}_{s}},\epsilon) \rangle &= f_{B^{*}_{s}}m_{B^{*}_{s}}\epsilon^{\mu}, \\
\nonumber
\langle0|\overline{s}\sigma^{\mu\nu}b|B^{*}_{s}(p_{B^{*}_{s}},\epsilon)\rangle &= -if^{T}_{B^{*}_{s}}(p^{\mu}_{B^{*}_{s}}\epsilon^{\nu}-\epsilon^{\mu}p^{\nu}_{B^{*}_{s}}), \\
\langle 0 | \overline{s}\gamma^\mu \gamma_5 b | B_s^* \rangle &= 0,
\label{fBsDefn}
\end{align}
where $\epsilon^{\mu}$ is the polarization vector of the $B_s^*$ meson. In the heavy quark limit, these are related to the decay constant of $B_s$ as,
$\langle 0 | \overline{s}\gamma^{\mu}\gamma_5 b |B_{s}(p)\rangle = - i f_{B_s} p^\mu$, and hence,
\begin{align} \nonumber
f_{B_s^*} &= f_{B_s} \bigg[1- \frac{2 \alpha_s}{3 \pi}\bigg]  \\ f_{B_s^*}^T &= f_{B_s}\bigg[1 + \frac{2 \alpha_s}{3 \pi}\Big(\log\Big(\frac{m_b}{\mu}\Big) - 1\Big)\bigg]
\end{align}
We use the relations in the heavy quark limit, $f_{B_s^*}/f_{B_s} = f_{B_s^*}^T/f_{B_s} = 0.953$ as given in \cite{Grinstein:2015aua}, which hold upto $O(\alpha_s)$. 
 The SM amplitude for $B_s^* \rightarrow \mu^+ \mu^-$ is given by,
 \begin{align} 
 \mathcal{M}^{SM} &= -\frac{\alpha G_F}{2\sqrt{2} \pi} V_{ts}^* V_{tb} f_{B_s^*} m_{B_s^*} \epsilon^{\mu} \bigg[(C_9^{eff} + 2 \frac{m_b f_{B_s^*}^T}{m_{B_s^*} f_{B_s^*} } C_7^{eff} (\overline{\mu} \gamma_{\mu} \mu) + C_{10 } (\overline{\mu} \gamma_{\mu} \gamma_{5} \mu) \bigg]
 \end{align}
and the decay rate is calculated to be \cite{Grinstein:2015aua}
\begin{align}
\Gamma(B^{*}_{s}\longrightarrow\mu^{+}\mu^{-})&=\frac{G^{2}_F\alpha^{2}}{96\pi^{3}}| V_{tb}V^{*}_{ts} |^{2} f^{2}_{B^{*}_{s}}m^{2}_{B^{*}_{s}}\sqrt{m^{2}_{B^{*}_{s}}-4m^{2}_{\mu}} \bigg[ \bigg| C^{eff}_{9}(m_{B_s^*}^2)+2\frac{m_{b}f_{B_s^*}^T}{m_{B^{*}_{s}}f_{B_s^*}}C^{eff}_{7}(m_{B_s^*}^2) \bigg|^{2}+ \bigg| C_{10} \bigg|^{2}\bigg]
\end{align} 
The decay rate shows explicit dependence on the Wilson Coefficients $C_7^{\mathrm{eff}}$ and $C_9^{\mathrm{eff}}$ and operators $O_{7,9}$ unlike the decay of pseudoscalar meson $B_s$. To obtain the numerical result for the decay rate, we use the values of the SM Wilson Coefficients upto NNLL accuracy as given in \cite{Altmannshofer:2008dz}. We use the values of other input parameters as follows: $\alpha_{em}= 1/127.94$, $f_{B_s^*}=0.2284 \pm 0.037\,\, \mathrm{GeV}$, $m_{B_s^*} = 5415.4 \pm 2.25\,\, \mathrm{MeV}$ and obtain the decay rate as,
\begin{align}
\Gamma(B_s^* \rightarrow \mu^+ \mu^-)|_\mathrm{SM} = 1.14 \pm 0.04 \times 10^{-18}\, \mathrm{GeV}.
\end{align}
To compute the branching ratio of $B_s^* \rightarrow \mu^+ \mu^- $, we need to know the total decay width of $B_s^*$ meson which is yet not known precisely from theoretical calculations or measurements. In order to get an estimate on the branching ratio, it is assumed that the total decay width of $B_s^*$, $\Gamma(B_{s^*}^{\mathrm{tot}})$ is comparable to the dominant decay process $B_s^* \rightarrow B_s \gamma$. From current experimental data and recent lattice QCD results, the decay width of $B_s^* \rightarrow B_s \gamma$ is found to be $\Gamma(B_s^* \rightarrow B_s \gamma)= 0.10 \pm 0.05 \,\, \mathrm{KeV}$ \cite{Grinstein:2015aua}. Using this, the branching ratio of $B_s^* \rightarrow \mu^+ \mu^-$ in the SM is calculated to be,
\begin{equation*}
BR(B^{*}_{s}\longrightarrow \mu^{+}\mu^{-})\vert_{SM}=
(1.14 \pm 0.57) \bigg(\frac{0.10 \pm 0.05\, \mathrm{KeV}}{\Gamma^{tot}_{B^{*}_{s}}}\bigg)\times 10^{-11}.
\end{equation*}
The SM branching ratio thus obtained for this process is roughly two orders of magnitude smaller than that of $B_s \rightarrow \mu^+ \mu^-$. 

\subsection{Branching ratio  of $B^{*}_{s} \rightarrow \mu^+\mu^-$ with new physics contributions}
\label{subsecB}
To study new physics effects in $B^{*}_{s} \rightarrow \mu^+\mu^-$ decay, we consider the addition of $V$, $A$, $S$ and $P$  operators to the SM effective Hamiltonian of $b \rightarrow s \mu^+ \mu^-$.  The effective Hamiltonian in the presence of these new physics operators is gives by,
\begin{eqnarray}
\mathcal{H}_{\mathrm{eff}}(b \rightarrow s \mu^+ \mu^-) = \mathcal{H}^{SM} + \mathcal{H}^{VA} + \mathcal{H}^{SP},
\end{eqnarray}
where $\mathcal{H}^{VA}$ and $\mathcal{H}^{SP}$ are as 
\begin{align} \nonumber
\mathcal{H}^{VA} &=  \frac{\alpha G_F}{\sqrt{2} \pi} V_{ts}^* V_{tb} \bigg[C_9^{NP} (\overline{s} \gamma^{\mu} P_L b)(\overline{\mu} \gamma_{\mu} \mu) + C_{10 }^{NP} (\overline{s} \gamma^{\mu} P_L b)(\overline{\mu} \gamma_{\mu} \gamma_{5} \mu)  \\
 &~~~~~~~~~~~~~~~ + C_9^{'NP}(\overline{s} \gamma^{\mu} P_R b)(\overline{\mu} \gamma_{\mu} \mu) + C_{10 }^{'NP} (\overline{s} \gamma^{\mu} P_R b)(\overline{\mu} \gamma_{\mu} \gamma_{5} \mu) \bigg]  \\  \nonumber
 \mathcal{H}^{SP} &=  \frac{\alpha G_F}{\sqrt{2} \pi} V_{ts}^* V_{tb} \bigg[R_S (\overline{s}  P_R b)(\overline{\mu} \mu) + R_P (\overline{s}  P_R b)(\overline{\mu} \gamma_{5} \mu)  \\
 &~~~~~~~~~~~~~~~ + R_S^{'}(\overline{s}  P_L b)(\overline{\mu} \mu) + R_P^{'} (\overline{s}  P_L b)(\overline{\mu} \gamma_{5} \mu) \bigg] 
\end{align}

where $C_9^{NP}, C_{10}^{NP},C_9^{'NP}, C_{10}^{'NP}, R_S, R_P, R_S^{'}, R_P^{'}$ are new physics couplings. 

We first compute the decay rate by considering new physics in the form of $S$ and $P$ operators. From the structure of the these operators, it can be seen that the matrix elements which appear in the calculation of the amplitude are $\langle 0 | \overline{s} b | B_s^* \rangle$ and $\langle 0 | \overline{s} \gamma^5 b | B_s^*\rangle$. Using the first and third relations defined in Eq.\eqref{fBsDefn}, one can show that,
\begin{align}
\langle 0 | \overline{s} b | B_s^* \rangle &= 0 , \\
\langle 0 | \overline{s} \gamma^5 b | B_s^* \rangle &= 0.
\end{align}
Hence the  branching ratio of $B_s^* \rightarrow \mu^+ \mu^-$ is not affected by new physics in the form of $S$ and $P$ operators. 

We now examine the contribution from $V$ and $A$ operators. Note that the matrix elements accompanying $C_9^{NP}$, and $C_9^{'NP}$ in the Hamiltonian for $V$ and $A$ contributions have the same Lorentz structure as the SM one for $C_9$, while the matrix elements accompanying $C_{10}^{NP}$, and $C_{10}^{'NP}$ are the same as the SM ones for $C_{10}$. The only difference being that unlike SM, new physics has right-handed chiral operator as well. Using the relationship between the matrix elements and decay constants  defined in Eq. \ref{fBsDefn}, the decay rate including NP VA contribution is obtained to be,
\begin{align*}
\Gamma(B^{*}_{s}\rightarrow\mu^{+}\mu^{-})&=\frac{G^{2}_F\alpha^{2}}{96\pi^{3}}\vert V_{tb}V^{*}_{ts}\vert^{2}f^{2}_{B^{*}_{s}}m^{2}_{B^{*}_{s}}\sqrt{m^{2}_{B^{*}_{s}}-4m^{2}_{\mu}}
\bigg[~\bigg|C^{eff}_{9}(m_{B_s^*}^2) +2 \frac{m_{b}f_{B_s^*}^T}{m_{B^{*}_{s}}f_{B_s^*}}C^{eff}_{7}(m_{B_s^*}^2) + C^{NP}_{9} \\& + C^{'NP}_{9}\bigg|^2 
+\bigg| C_{10}+ C^{NP}_{10}+C^{'NP}_{10}\bigg|^{2}~\bigg].
\end{align*}

\section{Methodology}

As new physics in the form of $S$ and $P$ operators do not contribute to the branching ratio of $B_s^* \to \mu^+ \mu^-$, we consider new physics only in the form of $V$ and $A$ operators. We consider various possible combinations of these new physics operators and obtain constraints on their coefficients by doing a global fit to all CP conserving observables in the  $b \to s\mu^{+} \mu^{-}$ sector. Most of these observables probe the kinematical distribution in $B\to K^* \mu^+\mu^-$ and $B^{0}_{s}\longrightarrow \phi \mu^{+}\mu^{-}$. The observables used in the fit are:

\begin{enumerate}

	\item The branching ratio of $B_{s} \to \mu^{+}\mu^{-}$ which is $(2.9 \pm0.7) \times 10^{-9}$ \cite{Aaij:2013aka,CMS:2014xfa}.

        \item The measurements of $R_K$ \cite{rk} and $R_{K^*}$ \cite{rkstar}.
	
	\item The differential branching ratio of $B^{0}\longrightarrow K^{0}\mu^{+}\mu^{-}$ \cite{Aaij:2014pli}.
	
	\item The differential branching ratio of $B^{+}\longrightarrow K^{+}\mu^{+}\mu^{-}$ \cite{Aaij:2014pli,CDFupdate}.

       \item The experimental measurements for the differential branching ratio of $B\rightarrow X_{s}\mu^{+}\mu^{-}$ \cite{Lees:2013nxa}.

	\item The nine measured observables in different $q^{2}$ bins in the decay $B^{0}\longrightarrow K^{*0}\mu^{+}\mu^{-}$ \cite{Kstarlhcb2}.

	\item The differential branching ratio of $B^{+}\longrightarrow K^{*+}\mu^{+}\mu^{-}$ \cite{Aaij:2016flj}.
	
	\item  The measurements of the angular observables and the differential branching ratio  of $B^{0}_{s}\longrightarrow \phi \mu^{+}\mu^{-}$ \cite{bsphilhc2,CDFupdate}.

 \end{enumerate}

A $\chi^2$ fit is done by using CERN minimization code {\tt MINUIT} \cite{James:1975dr,James:1994vla}. The  $\chi^2$  function is constructed as
\begin{equation}
\chi^2(C_i) = (\mathcal{O}_{th}(C_i) -\mathcal{O}_{exp})^T \, \mathcal{C}^{-1} \,
(\mathcal{O}_{th}(C_i) -\mathcal{O}_{exp})\,.
\end{equation}  
The $\chi^2$ function is minimized to get the best fit points and the theoretical predictions, $\mathcal{O}_{th}(C_i)$ are calculated using {\tt flavio} \cite{flavio}. $\mathcal{O}_{exp}$ are the experimental measurements of the
observables used in the fit.  We obtained the total covariance matrix $\mathcal{C}$  by adding the individual theoretical and experimental covariance
matrices.

We consider all possible combinations of new physics operators and obtain $\Delta\chi^2$ between the new physics best-fit points and SM best fit point. The fit results are presented Table \ref{pred}. We consider new physics only in the Wilson coefficients defined in Table \ref{pred} and all other Wilson coefficients are considered as SM like. We want to see if any new physics scenario can provide large enhancement in the branching ratio of $B_s^* \to \mu^+ \mu^-$ above its SM value.

\section{Results and Discussions}

The fit results for various new physics scenarios, along with the corresponding predictions for the branching ratio of $B_s^* \to \mu^+ \mu^-$, are presented in Table \ref{pred}.

\begin{table}[h]
	\begin{center}		\begin{tabular}{ | l | p{5cm} |p{1.5cm} |p{4cm} |}
			\hline Scenario & New physics couplings &$\Delta\chi^2$ & Branching Ratio\\ \hline
$C_i=0\,\,\rm (SM)$ & - & 0 & $(1.14 \pm 0.57)\times 10^{-11}$\\ \hline		
$C_9^{NP}$ &$-1.24\pm 0.18$  & $43.27$& $(0.86 \pm 0.43)\times 10^{-11}$\\ \hline 
$C_{10}^{NP}$ &$0.91\pm 0.19$  &$29.47$ & $(0.91 \pm 0.46)\times 10^{-11}$\\ \hline 
$C_9^{'}$  & $0.13\pm 0.16$ &$0.66$ & $(1.18 \pm 0.59)\times 10^{-11}$ \\ \hline 
$C_{10}^{'}$   & $-0.11\pm 0.13$ &$0.68$ & $(1.17 \pm 0.59)\times 10^{-11}$\\ \hline 
$C_9^{NP}=C_{10}^{NP}$    &$0.01\pm 0.18$  &$0.001$ &$(1.14 \pm 0.58)\times 10^{-11}$\\ \hline 
$C_9^{NP}=-C_{10}^{NP}$    &$-0.65 \pm 0.11$  & $43.04$&$(0.81 \pm 0.41)\times 10^{-11}$\\ \hline 
$C_9^{'}=C_{10}^{'}$    &$-0.04 \pm 0.17$  &$0.06$ &$(1.14 \pm 0.58)\times 10^{-11}$\\ \hline 
$C_9^{'}=-C_{10}^{'}$    &$0.07\pm 0.08$  & $0.81$&$(1.18 \pm 0.59)\times 10^{-11}$\\ \hline 
$[C_9^{NP},C_{10}^{NP}]$    &$[-1.10,0.33]$  & $47.33$&$(0.80 \pm 0.40)\times 10^{-11}$\\ \hline
$[C_9^{'},C_{10}^{'}]$    &$[0.08,-0.07]$  & $0.81$&$(1.18 \pm 0.60)\times 10^{-11}$\\ \hline
$[C_9^{NP}=C_{10}^{NP},C_9^{'}= C_{10}^{'}]$    &$[-0.02,-0.02]$  & $0.07$&$(1.15 \pm 0.58)\times 10^{-11}$\\  \hline
$[C_9^{NP}=-C_{10}^{NP},C_9^{'}= -C_{10}^{'}]$    &$[-0.67,0.16]$  & $46.27$&$(0.88 \pm 0.44 )\times 10^{-11}$\\  \hline
$[C_9^{NP},C_{10}^{NP},C_9^{'},C_{10}^{'}]$    &$[-1.31,0.26,0.34,-0.25]$  & $56.04$&$(0.91 \pm 0.48)\times 10^{-11}$\\  \hline
\end{tabular}
\caption{Calculation of the branching ratios of $B_s^* \to \mu^+ \mu^-$ for  various new physics scenarios. Here $\Delta\chi^2 = \chi^2_{\rm SM}-\chi^2_{\rm bf}$ and $\chi^2_{\rm bf}$ is the $\chi^2$ at the best fit points. We provide 1$\sigma$ range of the new physics couplings for the one parameter fits and the central values for multiple parameter fits.}
\label{pred}
\end{center}
\end{table}

It is obvious from Table \ref{pred} that none of the new physics scenarios can provide large enhancement in the branching ratio of $B_s^* \to \mu^+ \mu^-$ above its SM value. In scenarios where a good fit to the data is obtained, $Br(B^*_s \rightarrow \mu^+ \mu^-)$ is seen to be suppressed as compared to the SM value. Hence, most likely, the future measurements are expected to observe $B_s^* \to \mu^+ \mu^-$ decay with a branching ratio less than its SM prediction.

\section{Conclusions}
The new physics sensitivity of $B^*_s \rightarrow \mu^+ \mu^-$ decay is quite complementary to that of $B_s \rightarrow \mu^+ \mu^-$ decay as it is sensitive to different combinations of Wilson coefficients. More importantly, this decay is theoretically very clean. The decay rate can be accurately predicted in the standard model provided the $B^*_s$ decay width is well known.
In this work we perform a model independent analysis of new physics in $B^*_s \rightarrow \mu^+ \mu^-$ decay with an intend to identify the Lorentz structure of new physics  which can provide large enhancement in the branching ration of $B^*_s \rightarrow \mu^+ \mu^-$ above its SM value. For this, we consider new physics in the form of $V$, $A$, $S$ and $P$ operators. We show that the $S$ and $P$ operators do not contribute to $Br(B^*_s \rightarrow \mu^+ \mu^-)$. We then perform a global fit to all relevant $b \to s \mu^+ \mu^-$ data for different combinations of new physics $V$ and $A$ operators. For each of these scenarios, we predict $Br(B^*_s \rightarrow \mu^+ \mu^-)$. We find that none of these scenarios  can significantly enhance  $Br(B^*_s \rightarrow \mu^+ \mu^-)$.  All new physics operators which provide a good fit to the present $b \to s \mu^+ \mu^-$ data indicate suppression in $Br(B^*_s \rightarrow \mu^+ \mu^-)$ in comparison to its SM prediction. 

\section{Acknowledgment}
We thank Ashutosh Kumar Alok for useful discussions.


\begin{thebibliography}{10}

\bibitem{rk} 
  R.~Aaij {\it et al.} [LHCb Collaboration],
  Phys.\ Rev.\ Lett.\  {\bf 113}, 151601 (2014)
  [arXiv:1406.6482 [hep-ex]].

\bibitem{Hiller:2003js} 
  G.~Hiller and F.~Kruger,
  Phys.\ Rev.\ D {\bf 69}, 074020 (2004)
  [hep-ph/0310219].
  
  \bibitem{Bordone:2016gaq} 
  M.~Bordone, G.~Isidori and A.~Pattori,
  Eur.\ Phys.\ J.\ C {\bf 76}, no. 8, 440 (2016)
  [arXiv:1605.07633 [hep-ph]].
  
  
 \bibitem{rkstar} 
  R.~Aaij {\it et al.} [LHCb Collaboration],
  JHEP {\bf 1708}, 055 (2017)
  [arXiv:1705.05802 [hep-ex]].

\bibitem{Kstarlhcb1}
R.~Aaij {\it et al.} [LHCb Collaboration],
  Phys.\ Rev.\ Lett.\  {\bf 111}, 191801 (2013)
  [arXiv:1308.1707 [hep-ex]].

\bibitem{Kstarlhcb2}
R.~Aaij {\it et al.} [LHCb Collaboration],
  JHEP {\bf 1602}, 104 (2016)
  [arXiv:1512.04442 [hep-ex]].
 
\bibitem{KstarBelle}
A.~Abdesselam {\it et al.} [Belle Collaboration],
  arXiv:1604.04042 [hep-ex].

\bibitem{sm-angular} S.~Descotes-Genon, T.~Hurth, J.~Matias and J.~Virto,
  JHEP {\bf 1305}, 137 (2013)
  [arXiv:1303.5794 [hep-ph]].

\bibitem{kstaratlas}
ATLAS Collaboration, 
Tech.\ Rep.\ ATLAS-CONF-2017-023, CERN, Geneva, 2017.

\bibitem{kstarcms}
CMS Collaboration, 
in proton-proton collisions at $\sqrt{s} = 8$ TeV,'' 
Tech.\ Rep.\ CMS-PAS-BPH-15-008, CERN, Geneva, 2017.

\bibitem{bsphilhc1}
R.~Aaij {\it et al.} [LHCb Collaboration],
  JHEP {\bf 1307}, 084 (2013)
  [arXiv:1305.2168 [hep-ex]].

\bibitem{bsphilhc2}
R.~Aaij {\it et al.} [LHCb Collaboration],
  JHEP {\bf 1509}, 179 (2015)
  [arXiv:1506.08777 [hep-ex]].
  
    \bibitem{Capdevila:2017bsm} 
  B.~Capdevila, A.~Crivellin, S.~Descotes-Genon, J.~Matias and J.~Virto,
  arXiv:1704.05340 [hep-ph].
  
  
  \bibitem{Alok:2010zd} 
  A.~K.~Alok, A.~Datta, A.~Dighe, M.~Duraisamy, D.~Ghosh and D.~London,
  JHEP {\bf 1111}, 121 (2011)
  [arXiv:1008.2367 [hep-ph]].
  
  \bibitem{Alok:2011gv} 
  A.~K.~Alok, A.~Datta, A.~Dighe, M.~Duraisamy, D.~Ghosh and D.~London,
  JHEP {\bf 1111}, 122 (2011)
  [arXiv:1103.5344 [hep-ph]].
  
\bibitem{Descotes-Genon:2013wba} 
  S.~Descotes-Genon, J.~Matias and J.~Virto,
  Phys.\ Rev.\ D {\bf 88}, 074002 (2013)
  [arXiv:1307.5683 [hep-ph]].
  
  \bibitem{Altmannshofer:2014rta} 
  W.~Altmannshofer and D.~M.~Straub,
  Eur.\ Phys.\ J.\ C {\bf 75}, no. 8, 382 (2015)
  [arXiv:1411.3161 [hep-ph]].
  
  \bibitem{Altmannshofer:2015sma} 
  W.~Altmannshofer and D.~M.~Straub,
  arXiv:1503.06199 [hep-ph].
 
  \bibitem{Hurth:2016fbr} 
  T.~Hurth, F.~Mahmoudi and S.~Neshatpour,
  Nucl.\ Phys.\ B {\bf 909}, 737 (2016)
  [arXiv:1603.00865 [hep-ph]].
  
  \bibitem{Capdevila:2016ivx} 
  B.~Capdevila, S.~Descotes-Genon, J.~Matias and J.~Virto,
  JHEP {\bf 1610}, 075 (2016)
  [arXiv:1605.03156 [hep-ph]].

  \bibitem{Altmannshofer:2017fio} 
  W.~Altmannshofer, C.~Niehoff, P.~Stangl and D.~M.~Straub,
  Eur.\ Phys.\ J.\ C {\bf 77}, no. 6, 377 (2017)
  [arXiv:1703.09189 [hep-ph]].
  
  \bibitem{Alok:2017jgr} 
  A.~K.~Alok, B.~Bhattacharya, D.~Kumar, J.~Kumar, D.~London and S.~U.~Sankar,
  Phys.\ Rev.\ D {\bf 96}, no. 1, 015034 (2017)
  [arXiv:1703.09247 [hep-ph]].
  
  
  \bibitem{Alok:2017sui} 
  A.~K.~Alok, B.~Bhattacharya, A.~Datta, D.~Kumar, J.~Kumar and D.~London,
  arXiv:1704.07397 [hep-ph].
  
  
\bibitem{Grinstein:2015aua} 
  B.~Grinstein and J.~Martin Camalich,
  Phys.\ Rev.\ Lett.\  {\bf 116}, no. 14, 141801 (2016)
  [arXiv:1509.05049 [hep-ph]].

   \bibitem{Xu:2015eev} 
  G.~Z.~Xu, Y.~Qiu, C.~P.~Shen and Y.~J.~Zhang,
  Eur.\ Phys.\ J.\ C {\bf 76}, no. 11, 583 (2016)
  [arXiv:1601.03386 [hep-ph]].
  
  \bibitem{Sahoo:2016edx} 
  S.~Sahoo and R.~Mohanta,
  J.\ Phys.\ G {\bf 44}, no. 3, 035001 (2017)
  [arXiv:1612.02543 [hep-ph]].
  
  \bibitem{Altmannshofer:2008dz}
  A.~Wolfgang, P.~Ball, A.~Bharucha, A.~J.~Buras, et.al.,
  JHEP {\bf 0901}, 091 (2009),
  [arXiv:0811.1214 [hep-ph]]

\bibitem{Aaij:2013aka}
R.~Aaij {\it et al.} [LHCb Collaboration],
  Phys.\ Rev.\ Lett.\  {\bf 111}, 101805 (2013)
  [arXiv:1307.5024 [hep-ex]].

\bibitem{CMS:2014xfa}
V.~Khachatryan {\it et al.} [CMS and LHCb Collaborations],
  Nature {\bf 522}, 68 (2015)
  [arXiv:1411.4413 [hep-ex]].

\bibitem{Aaij:2014pli}
R.~Aaij {\it et al.} [LHCb Collaboration],
  JHEP {\bf 1406}, 133 (2014)
  [arXiv:1403.8044 [hep-ex]].


\bibitem{CDFupdate}
\textbf{CDF} Collaboration, 
  CDF public note 10894.

 \bibitem{Lees:2013nxa}
J.~P.~Lees {\it et al.} [BaBar Collaboration],
  Phys.\ Rev.\ Lett.\  {\bf 112}, 211802 (2014)
  [arXiv:1312.5364 [hep-ex]].



\bibitem{Aaij:2016flj}
  R.~Aaij {\it et al.} [LHCb Collaboration],
  JHEP {\bf 1611}, 047 (2016)
  [arXiv:1606.04731 [hep-ex]].

 \bibitem{James:1975dr}
  F.~James and M.~Roos,
  Comput.\ Phys.\ Commun.\  {\bf 10}, 343 (1975).

\bibitem{James:1994vla}
  F.~James,
  CERN-D-506, CERN-D506.
  
\bibitem{flavio}
David Straub, \textit{flavio v0.11, 2016.}
  \href{http://dx.doi.org/10.5281/zenodo.59840}{http://dx.doi.org/10.5281/zenodo.59840}
  

  

  
  

  
\end{thebibliography}
\end{document}